\documentclass[twocolumn,nofootinbib,aps,pre,superscriptaddress]{revtex4-2}
\pdfoutput=1

\usepackage{graphicx}
\usepackage{dcolumn}
\usepackage{bbold}
\usepackage{csquotes}
\usepackage{amsmath}
\usepackage{xcolor}
\usepackage{bm}
\usepackage{hyperref}
\usepackage[symbol]{footmisc}

\usepackage{physics} 
\usepackage[toc,page]{appendix}

\begin{document}



\title{Quantum targeted energy transfer through machine learning tools}

\author{I. Andronis}
\email{ph4783@edu.physics.uoc.gr}
\affiliation{Department of Physics, University of Crete, Heraklion 70013, Greece}

\author{G. Arapantonis}
\email{garapan1@jh.edu}
\affiliation{Department of Physics, University of Crete, Heraklion 70013, Greece}
\affiliation{William H. Miller III Department of Physics \& Astronomy, Johns Hopkins University, Baltimore, Maryland 21218, USA}
\author{G. D. Barmparis}
\affiliation{Department of Physics, University of Crete, Heraklion 70013, Greece}
\author{G. P. Tsironis}
\affiliation{Department of Physics, University of Crete, Heraklion 70013, Greece}
\affiliation{John A. Paulson School of Engineering and Applied Sciences, Harvard University, Cambridge, Massachusetts 02138, USA}

\date{\today}

\begin{abstract}
In quantum targeted energy transfer, bosons are transferred from a certain crystal site to an alternative one, utilizing a nonlinear resonance configuration similar to the classical targeted energy transfer.\@ We use a novel computational method based on machine learning algorithms in order to investigate selectivity as well as efficiency of the quantum transfer in the context of a dimer and a trimer system.\@ We find that our method identifies resonant quantum transfer paths that allow boson transfer in unison.\@ The method is readily extensible to larger lattice systems involving nonlinear resonances. 
\end{abstract}

\maketitle

\section{Introduction}
\label{Section:Intro}
Nonlinear dynamical systems are notoriously difficult to study analytically and in most cases, one needs to resort to numerical methods for their analysis~\cite{ott_2002}.\@ In the classical realm, they are typically described mathematically through coupled nonlinear differential equations that very scarcely admit exact solutions.\@ In this case, one resorts to direct numerical integration of the equations of motion.\@ In the quantum domain, on the other hand, even though the equations are linear, one needs to engage a very large part of the Hilbert space in order to find good, yet approximate, solutions~\cite{Akulin2014}.\@ Direct numerical methods, either in the classical or quantum domain, are relatively straightforward, yet might fail in large or strongly coupled systems.\@ The recent widespread of Machine Learning (ML) techniques and their implementation in the domain of dynamical systems aim to
both facilitate and also improve the discovery process in these systems~\cite{Chimeras}.\@ The aim of the present work is to use techniques motivated by ML and obtain results that would be otherwise very complex to derive.\@ The specific model we work with is that of Targeted Energy Transfer (TET) that was inspired by energy transfer processes in chlorophyll~\cite{Kopidakis2001}.\@ We have two targets here;  the first one is to show how ML-motivated techniques may be superior to standard numerical methods when applied in  quantum complex systems and the second is to find explicit results that would be otherwise much more difficult to obtain.  

In the semiclassical TET model, we focus on resonant  exciton transfer between non-identical molecules~\cite{Kopidakis2001}.\@ In each molecule a single energy state participates in the process, thus we have non-identical energy states coupled together via a non-zero transfer matrix element. In the simplest case of two molecules, we deal with a  non-degenerate linear dimer system.\@  Due to the energy mismatch, the exciton transfer from the first site to the second is non-resonant and thus occurs only partially.\@ In order to make the transfer resonant we need to add local interaction with additional degrees of freedom, such as phonons.\@ In the anti-adiabatic approximation, this procedure introduces effective qubic nonlinearities~\cite{HENNIG1999333}.\@ In TET, complete resonant transfer is restored for specific nonlinearity parameter configurations linking the local interaction with anti-adiabatic phonons and the actual energy mismatch.\@ The analysis is done within the context of the discrete nonlinear 
Schr{\"o}dinger (DNLS) equation,  a ubiquitous model for a large class of nonlinear phenomena~\cite{Eilbeck1985},\cite{Kenkre2022}.\@ A number of analytical results known for DNLS dimers are particularly useful for ML implementations in nonlinear systems~\cite{Kenkre1986, Tsironis1988,Tsironis1993}.

While in the case of the semiclassical TET dimer system the resonant transfer regime can be found analytically, similar analysis in larger systems is an arduous task.\@ In order to bypass this difficulty, ML-based approaches have been introduced that enable the discovery of nonlinear resonances in a straightforward way ~\cite{Barmparis2021,Tsironis2022}. The method was tested both in the TET dimer and also in other analytically known DNLS equation results.\@ Also, it was applied to the case of a TET trimer model.\@ This ML approach found readily the trimer resonances that were very difficult to obtain differently.\@ The implementation of ML in this semiclassical regime shows that its application can be very beneficial.

The next challenge is that of addressing the fully quantum TET regime; this is the aim of the present article. When the TET dimer is quantized with bosonic degrees of freedom a more general 
resonant condition, which involves the number of quanta, arises~\cite{Maniadis2004}.\@ When at resonance, these bosons may transfer collectively from the first site to the second in a way similar to the
semiclassical TET, although with rates depending on the boson number and the energy difference.\@ As in the semiclassical case, in the fully quantum TET, the resonant condition can be found analytically in the dimer model
\cite{Maniadis2004}, but any extension to larger systems is prohibitive analytically and involves a high computational cost.\@ We show in the present work that the implementation of ML methods can help in overcoming these difficulties and be able to obtain readily the required resonant transfer properties.

The structure of the present article is thus the following.\@ In the next section II, we introduce a general form of the DNLS equation that upon quantization leads to fully quantum TET models.\@ Once we 
define clearly the problem in the quantum case, we discuss in the following section III the ML technique we use.\@ More specifically, we discuss the choice of the Loss Function (LF) that enables the analysis of the resonant 
transfer both in the dimer and also more generally to arbitrary chains.\@ In section IV we detail the optimization method, and how the quantum resonant paths are found for different boson numbers. Section V is the central section of the article; we not only recover the exact dimer results but also apply the method to a trimer configuration.\@ This gives not only the new result of the resonant paths but also shows that the method is fully able to investigate in detail the specifics of the resonant transfer.\@  
Finally, in section VI we conclude and summarize the findings of the work and comment on possible extensions.
\section{From the semiclassical to the fully quantum TET}

It is known that the semiclassical DNLS equation can be derived from a classical Hamiltonian through the use
of Hamilton's equations~\cite{HENNIG1999333}.\@ This classical DNLS Hamiltonian is 

\begin{equation}{\label{classical_dnls_hamiltonian}}
\hspace*{-0.37cm}
  H = \sum_{k=1}^f \omega_k|\psi_k|^2 + \frac{1}{2}\chi_k|\psi_k|^4 - \lambda \sum_{k=1}^{f-1}(\psi^*_{k}\psi_{k+1} + \psi^*_{k+1}\psi_{k}),
\end{equation}
where $\omega_k$ and $\chi_k$ denote the frequency and nonlinearity parameter of the oscillator at site $k$ respectively.\@ Also, ($\psi^{*}_k,i\psi_k)$ is a pair of conjugate variables, and the parameter  $\lambda$ is the coupling  among neighboring sites.\@ For simplicity, we assume that it is the same between every pair of adjacent oscillators.\@ In order to move to the quantum mechanical case we need to focus on the Bose-Hubbard operator~\cite{Gersch1963}.\@ 
The DNLS model for the quantum domain may be seen to arise also from the Bose-Hubbard model by using the time-dependent variational principle~\cite{Amico1998}.\@ A simple way to quantize the Hamiltonian of Eq.~\eqref{classical_dnls_hamiltonian} is by substituting $\psi^*_k,\space \psi_k$ with the creation and annihilation operators $a^{\dagger}_k,\space a_k$ respectively, as expressed in the second quantization formalism~\cite{Pearsall2020}.\@ These operators obey to the commutation relations $[a_k, a^{\dagger}_m] = \delta_{km}$, and $[a_k, a_m]=0 $, where $\delta_{km}$ is the Kronecker delta.\@ Therefore, the Hamiltonian operator in the quantum case becomes
\begin{equation}{\label{bosehubbard_hamiltonian}}
    \hspace{-0.45cm}
  \hat{H} = \sum_{k=1}^f \omega_k\hat{N}_k + \frac{1}{2}\chi_k\hat{N}_k^2  -  \lambda\sum_{k=1}^{f-1}(\hat{a}^{\dagger}_{k}\hat{a}_{k+1} + \hat{a}^{\dagger}_{k+1}\hat{a}_{k}),
\end{equation}
where $\hat{N}_k = \hat{a}^{\dagger}_k \hat{a}_k$ is the boson number operator for the site $k$.\@ The dimension of the Hilbert space $\mathcal{H}_N$ for this problem is finite.\@ Each state corresponds to an allowed configuration of $N$ indistinguishable bosons occupying $f$ distinguishable sites-nonlinear oscillators, with repetitions.\@ Thus the dimension of the Hilbert space is
\begin{equation*}{\label{dim}}
  \mathcal{D} = \frac{(N+f-1)!}{N!(f-1)!}.
\end{equation*}

A basis associated with that problem is the one composed of the so-called Fock states $\ket{n} \equiv \ket{n_1, n_2, \dots, n_{f}}$, where $n_1, n_2, \dots, n_f$ is the number of bosons at each respective  site $1,\space 2,\dots,\space f$ at the state indexed as n. \hspace{-0.18cm}We discuss  the procedure of labeling the Fock states in section~\ref{Section:LF}. The occupation numbers $\{n_i\}$ are restricted to $\sum_i n_i = N$, while $n_i = 0,1,\dots,N$.\@ Additionally,\@ the Fock states are orthonormal, meaning
\begin{equation*}\label{innerprod}
  \bra{n}\ket{m} = \delta_{n_1m_1} \dots \delta_{n_fm_f}.
\end{equation*}

Continuing, the actions of the operators $\hat{a}_k,\hat{a}^{\dagger}_k,\hat{N}_k$ on each component $\ket{n}$ of the basis are described by
\begin{subequations}{\label{actions}}
  \begin{equation}\label{annihilation_act}
    \hat{a}_k \ket{\dots, n_k, \dots} = \sqrt{n_k} \ket{\dots,n_k-1, \dots},
  \end{equation}
  \begin{equation}\label{creation_act}
    \hat{a}^{\dagger}_k \ket{\dots, n_k, \dots} = \sqrt{n_k+1} \ket{\dots, n_{k}+1, \dots},
  \end{equation}
  \begin{equation}\label{number_act}
    \hat{N}_k \ket{\dots, n_k, \dots} = n_k \ket{\dots, n_{k}, \dots}.
  \end{equation}
\end{subequations}

\section{Determination of an appropriate Loss Function}
\label{Section:LF}

We now focus on the choice of an appropriate loss function.\@ We assume that the  \textit{donor} site has the lowest energy  with nonlinearity parameter $\chi_D$ while the highest energy site is the \textit{acceptor} site with nonlinearity parameter $\chi_A$. 
We further assume that all the bosons are placed initially (at time $t=0$) to the donor, and investigate the TET configurations that allow the complete transfer of these bosons to the acceptor site.\@ To achieve this, we employ an algorithm relying on the same principle as the ones presented in~\cite{Karniadakis2021,Barmparis2021}, where an optimization algorithm is used to minimize a quantity that is defined as the LF. The LF is usually associated with some physical parameters and thus, the problem becomes one where the algorithm, has to tune the parameters. 

The first step towards defining the LF is to construct a numerical scheme for calculating the matrix elements of the Hamiltonian in Eq.~\eqref{bosehubbard_hamiltonian}.\@ While this is usually a simple process, there are special intricacies, with the main one being the labeling of the Fock states.\@ We manage to overcome this obstacle by implementing a similar technique to~\cite{Zhang2010}, where the authors rank the states in \textit{lexicographic} order and assign indices $1, \space 2,\dots,\space \mathcal{D}$ to each configuration of bosons among the sites.\@ For instance, assuming $N=2$ and $f=3$, state 1 corresponds to $\ket{1} = \ket{2,0,0}$, state 2 corresponds to $\ket{2} = \ket{1,1,0}$ and so on, assigning every state-configuration to a distinct index.

Under this indexing policy, it is now straightforward to compute the elements $\hat{H}_{ij}$ similarly to~\cite{Maniadis2004}, considering that
\begin{subequations}
    \begin{equation}{\label{sum_terms}}
        \hat{H}_{ij} = \bra{i} \hat{H} \ket{j} \overset{(\ref{bosehubbard_hamiltonian})}{=} T_1 + T_2,
    \end{equation}
    \begin{equation}\label{hamiltonian_element1}
        T_1 \equiv \bra{i}\sum_{k=1}^{f} \left[\omega_k\hat{N_k} + \frac{1}{2}\chi_k(\hat{N_k})^2\right] \ket{j},
    \end{equation}
    \begin{equation}\label{hamiltonian_element2}
    T_2 \equiv -\lambda\bra{i} \sum_{k=1}^{f-1} \left[\hat{a}^{\dagger}_{k}\hat{a}_{k+1} + \hat{a}^{\dagger}_{k+1}\hat{a}_{k}\right] \ket{j}.
    \end{equation}
\end{subequations}
Calculating $T_1$ is simple since it involves the simple action of the boson number operator on state $\ket{j}$, as shown in Eq.~\eqref{number_act}.\@ That said, Eq.~\eqref{hamiltonian_element1} is equivalent to
\begin{eqnarray*}\label{he_term1_calc}
    T_1 =\sum_{k=1}^{f} \left[\omega_k j_k + \frac{1}{2}\chi_k(j_k)^2 \right] \delta_{ij},
\end{eqnarray*}
where $j_k$ stands for the number of bosons on site $k$ for the state $\ket{j}$.\@ Evaluating the second term is nontrivial because it involves the action of the operators $\hat{a}^\dagger_k\hat{a}_{k+1}$, $\hat{a}^{\dagger}_{k+1}\hat{a}_{k}$.\@ The consecutive action of these operators can be explored by referring to Eq.~\eqref{annihilation_act} and Eq.~\eqref{creation_act}.\@ The operators $\hat{a}^\dagger_k$, $\hat{a}_k$ create and annihilate a boson at a given site $k$ respectively. However, when a pair of operators like $\hat{a}^\dagger_k\hat{a}_{k+1}$ (or $\hat{a}^{\dagger}_{k+1}\hat{a}_{k}$) acts on a Fock state $\ket{j}$, it creates a boson at the site $k$ (or $k+1$) but also destroys a boson at the site $k+1$ (or $k$).\@ Thus, their action on a Fock state conserves the total number of bosons and the resulting state is going to be, up to a constant, another state in $\mathcal{H}_N$. Specifically,
\begin{subequations}{\label{cross_terms}}
    \begin{equation*}\label{create_annihilate_act}
        \bra{i}\hat{a}^{\dagger}_{k+1}\hat{a}_{k}\ket{j} = \sqrt{j_k(j_{k+1}+1)} \delta_{ip}  \equiv C_k^{(p)}\delta_{ip},
    \end{equation*}
    \begin{equation*}\label{annihilate_create_act}
        \bra{i}a_k^\dagger a_{k+1}\ket{j} =  \sqrt{j_{k+1}(j_k+1)} \delta_{im}  \equiv D_k^{(m)}\delta_{im},
    \end{equation*}
\end{subequations}
where the two new states $\ket{p}, \ket{m}$ are
\begin{subequations}
\begin{equation*}{\label{new_state1}}
    \ket{p} = \ket{j_1,\dots,j_k-1,j_{k+1}+1 \dots,j_f},
\end{equation*}
\begin{equation*}{\label{new_state2}}
    \ket{m} = \ket{j_1,\dots,j_k+1,j_{k+1}-1 \dots,j_f}.
\end{equation*}
\end{subequations}
Combining the above yields
\begin{eqnarray*}{\label{he_term2_calc}}
  T_2 =  -\lambda \sum_{k=1}^{f-1} \ \left[ C_k^{(p)}\delta_{ip} +D_k^{(m)}\delta_{im} \ \right].
\end{eqnarray*}

\begin{figure}[t]
    \includegraphics{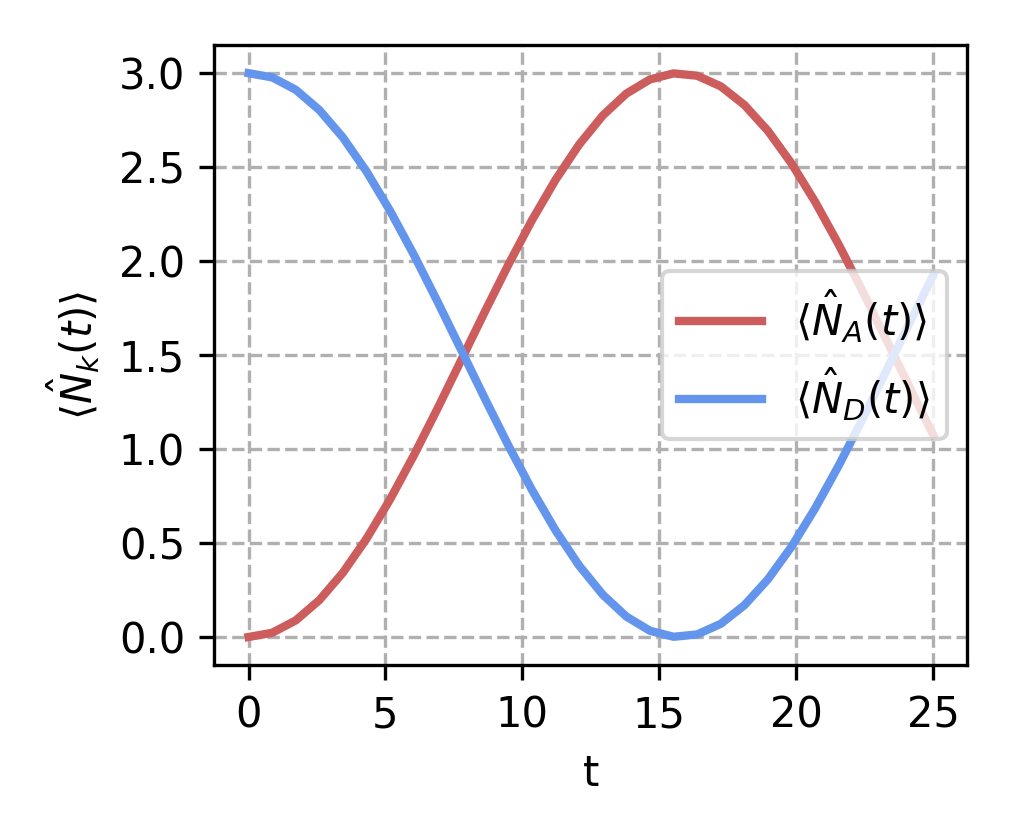}
    \caption{Time evolution of the expectation value of the number of bosons for the two sites of the dimer under the parameters ($\chi_A$, $\chi_D$, $\omega_A$, $\omega_D$, $\lambda$, $N$, \texttt{maxt}) = (-2,2, 3, -3, 0.1, 3, 25).\@ The blue line denotes the donor's expectation value, while the red one the acceptor's.} 
    \label{TimeEvolutionPlot}
\end{figure}
The matrix representation of the Hamiltonian can be produced and subsequently, the eigenstates and eigenvalues can be calculated. Everything is developed in \textit{Python}, using the \textit{Tensorflow}~\cite{tensorflow2015-whitepaper} library.

The initial distribution of bosons $\ket{\Psi(0)}$ can be expanded to the basis of the eigenstates $\ket{\psi_i}$
\begin{equation*}\label{initial_state_expansion}
  \ket{\Psi(0)} = \sum_{i=1}^{\mathcal{D}} C_{i} \ket{\psi_i},\ \ C_{i} = \bra{\psi_i} \ket{\Psi(0)}.
\end{equation*}
Also, these eigenstates can be expanded to the basis of the Fock states
\begin{equation*}\label{eigenstate_expansion}
  \ket{\psi_i} = \sum_{j=1}^{\mathcal{D}} b_{j,i} \ket{j},\ \ b_{j,i} = \bra{j} \ket{\psi_i}.
\end{equation*}
We can now express the time evolution of the initial distribution $\ket{\Psi(0)}$ by applying the time evolution operator $\hat{U}(t) = e^{-\textbf{i}\hat{H}t}:$
\begin{equation}\label{general_state_expansion}
\ket{\Psi(t)} =e^{-\textbf{i}\hat{H}t}\ket{\Psi(0)}
 = \sum_{i,j}^{\mathcal{D}} C_i b_{j,i} e^{-\textbf{i}E_it} \ket{j},    
\end{equation}
where $E_i$ is the $i$-th eigenvalue and $\textbf{i}$ is the imaginary unit.\@ Similarly, the time evolution of the average number of bosons at the site k is given by 
\begin{equation}{\label{avg_ksite}}
\langle{}\hat{N}_k(t)\rangle{} \ \ = \ \ \bra{\Psi(t)} \hat{N_k} \ket{\Psi(t)}.
\end{equation}
Combining Eq.~\eqref{general_state_expansion} and Eq.~\eqref{avg_ksite}, $\langle \hat{N}_k \rangle$ can be assessed in the following way
\begin{eqnarray}\label{avg_n}
\hspace{-0.4cm}
  \langle \hat{N}_k(t) \rangle= \sum_{i,j,n}^{\mathcal{D}} j_k C_n^*  C_i b^*_{j,n} b_{j,i} e^{\textbf{i}(E_n-E_i)t}.
\end{eqnarray}
We use Eq.~\eqref{avg_n} in order to compute the LF for the quantum TET problem.\@ Specifically, we time-evolve Eq.~\eqref{avg_n} for the acceptor energy level until a pre-defined time \verb{maxt{.\@ During this period, the oscillator of the $f^{th}$ site (acceptor) has completed a few oscillations.\@ The next step in computing the LF is to extract the maximum value from that time evolution.\@ Concluding, the LF is defined as
\begin{equation}{\label{Eq:LF}}
    LF = N- max\{ \langle \hat{N}_f(t) \rangle \} =N-max\{ \langle \hat{N}_A(t) \rangle\}.  
\end{equation}
In Fig.~\ref{TimeEvolutionPlot} one can observe the characteristic oscillatory behavior, in the dimer system, when the complete transfer occurs.
\par
In the present work, we  fix the frequency of the oscillators and optimize for the nonlinearity parameters of the oscillators.\@ While the LF might appear not to have any explicit connection to the  nonlinearity parameters we want to optimize for, we can use Tensorflow's $GradientTape$ to compute the derivatives with respect to these parameters.\@ In that way, we keep track of gradient information in terms of the trainable variables throughout the whole process described in this section;\@ from creating the Hamiltonian to calculating $\langle \hat{N}_k(t) \rangle$.\@ Using this information, an optimizer like $ADAM$ \cite{Kingma2014} can now update the parameters accordingly, so that the LF is minimized, signifying complete transfer.\@ It is important to note that, while many other optimization algorithms (simulated annealing \cite{Bertsimas1993},\@ particle swarm \cite{Kennedy1995},\@ differential evolution \cite{Brest2006, Elsayed2011}) were tried on this problem, we were not able to produce adequate results with none of them.

The parameter \verb{maxt{ is of major importance, because of the oscillation of the bosons between the sites.\@ If it is not set large enough, TET can be missed since the system would not have time to complete an oscillation, while it also has to be small enough, so that precious computational time is saved.\@ Thus, it has to be large enough to obtain at least one complete oscillation, producing this way essential information.\@ It is also observed that the period of each oscillator is proportional to $\lambda^{-1}$, so as one would expect, changes to these parameters should be made concurrently.
\section{Optimization of the Loss Function}
\begin{figure}[t]
    \includegraphics[scale=0.28]{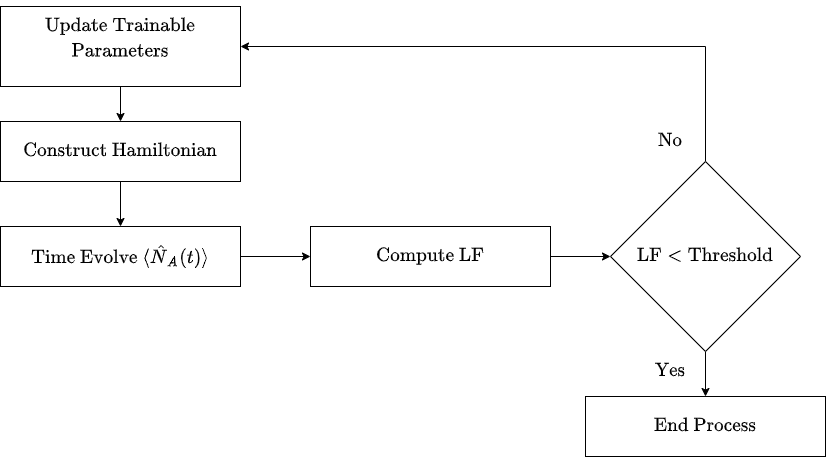}
    \caption{Schematic representation of the parameter optimization procedure.}
    \label{Training}
\end{figure}

Once we define the LF we may now proceed with its minimization through the procedure outlined in Fig.~\ref{Training}. We  emphasize the first step of this process graph, i.e.\@ a proper  update method, which is essential for the
present work.\@ We noticed by making several computational runs that optimizers did not work very well when initialized with a  random set of  parameters; this is due to the extreme selectivity of the TET resonant condition. Specifically, the parameter space that the LF maps to, exhibits slowly varying gradients everywhere, except for areas close to the optimal parameters for transfer, where relatively large gradients are present.

Our approach for bypassing this problem relies on exploring a wide range of the parameter space at the same time.\@ To be more specific, the optimization procedure is comprised of two distinct phases.\@ First, many \textit{test optimizers} are assigned some initial guesses and left to run simultaneously,\@ as different processes,\@ until a stopping condition is reached.\@ Then we gather the LF and the parameters that each test optimizer yielded at the last iteration.\@ For the second part of the optimization procedure, a \textit{main optimizer} is employed.\@ Its initial guesses are the parameters of the test optimizer with the best performance (i.e smallest LF). The purpose of this additional step is to further minimize the LF, if possible.

\begin{figure*}[t!]
    \includegraphics[scale=1]{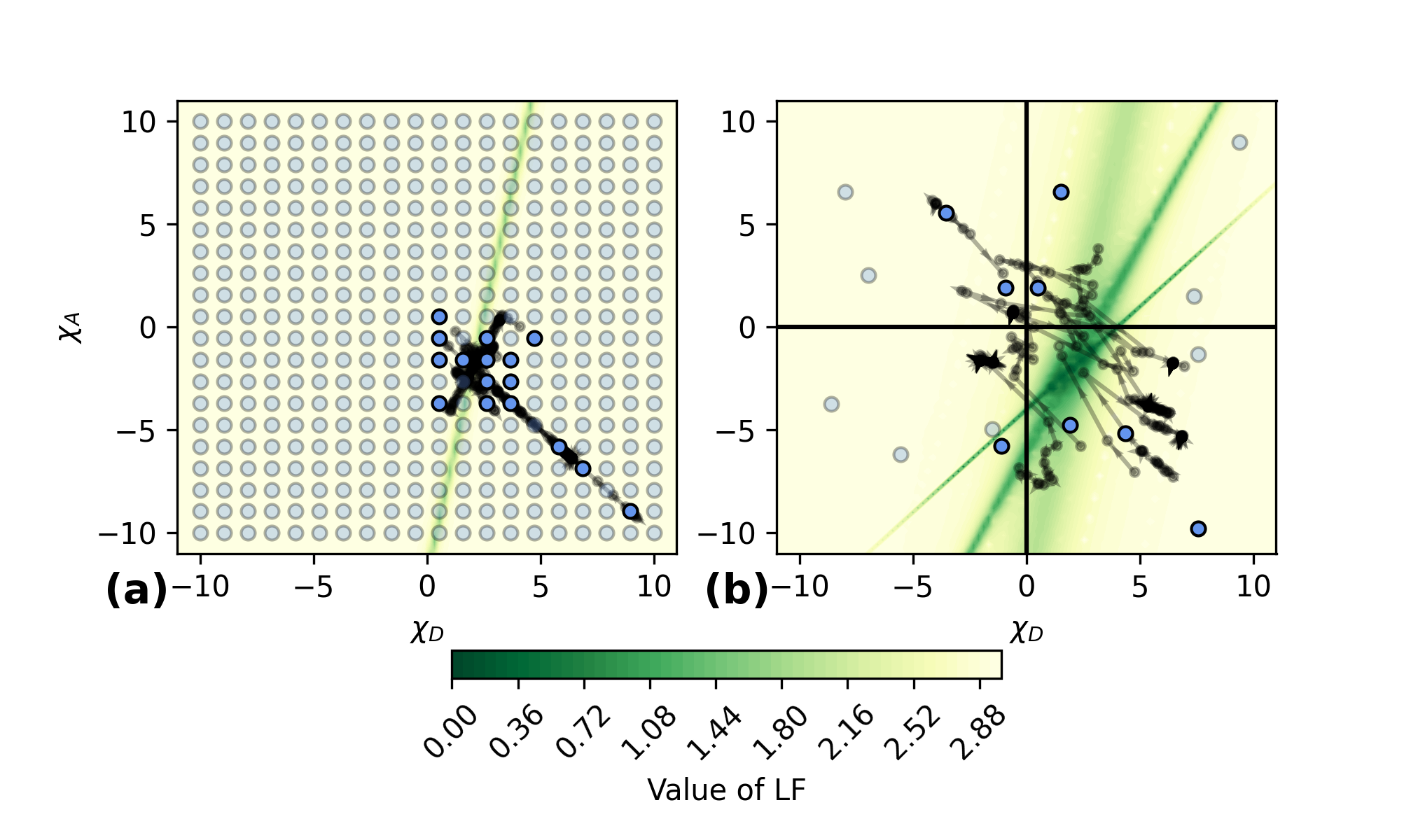}
    \caption{Defining the initial guesses of the test optimizers.\@ \textbf{(a)} An example using the \textit{grid} method, where we define 400 different initial guesses for the same number of test optimizers.\@ \textbf{(b)} Representative example of the second method, where splitting of the parameter space into regions is employed.\@ The black lines represent the boundaries among the four different regions from which 16 different guesses are sampled.\@ In both graphs, the test optimizers that were able to derive a $\text{LF} < 1$ are displayed with normal opacity and their trajectories, while the rest are faded.\@ Both figures refer to a dimer system with ($N$, $\omega_A$, $\omega_D$, \texttt{maxt}) = (3, 3, -3, 25). For the purposes of explaining the difference between the two methods, the system presented in the left figure has a coupling parameter of $\lambda = 0.1$, while for the system on the right figure $\lambda = 1$.}
    \label{Fig:Define_Initial_Guesses}
\end{figure*}

Two methods were developed for defining the initial guesses of the test optimizers, given that we choose the range of the parameter space they will investigate.\@ In the first approach, which is represented by Fig.~\ref{Fig:Define_Initial_Guesses}a, we define a \textit{grid} of points in the parameter space and use the lattice points as initial guesses for the test optimizers.\@ In the second method, we split the parameter space into a number of regions and chose random initial combinations of parameters from each region.\@ On the one hand, the former method has a high computational cost, since it requires a large number of optimizers running at the same time, but has a greater potential to derive the optimal parameters.\@ On the other hand, the latter is fast but less robust.\@ It is useful in cases of stronger coupling, as the gradients of the parameter space smooth out.\@ If the first run of any method does not produce favorable results we have the option of redefining the limits of the parameter space around the best parameters provided by the test optimizers. Nevertheless, choosing one of the above methodologies relies on the problem at hand.\@ In our case, both of them produce accurate results.\@ Moreover, we need to disambiguate that the main optimizer is an optional step, that aims to corroborate TET, by producing a LF lower than the one deduced from the test optimizers.

\begin{figure*}[ht!]
    \includegraphics[width=15.5cm,height=10.4cm]{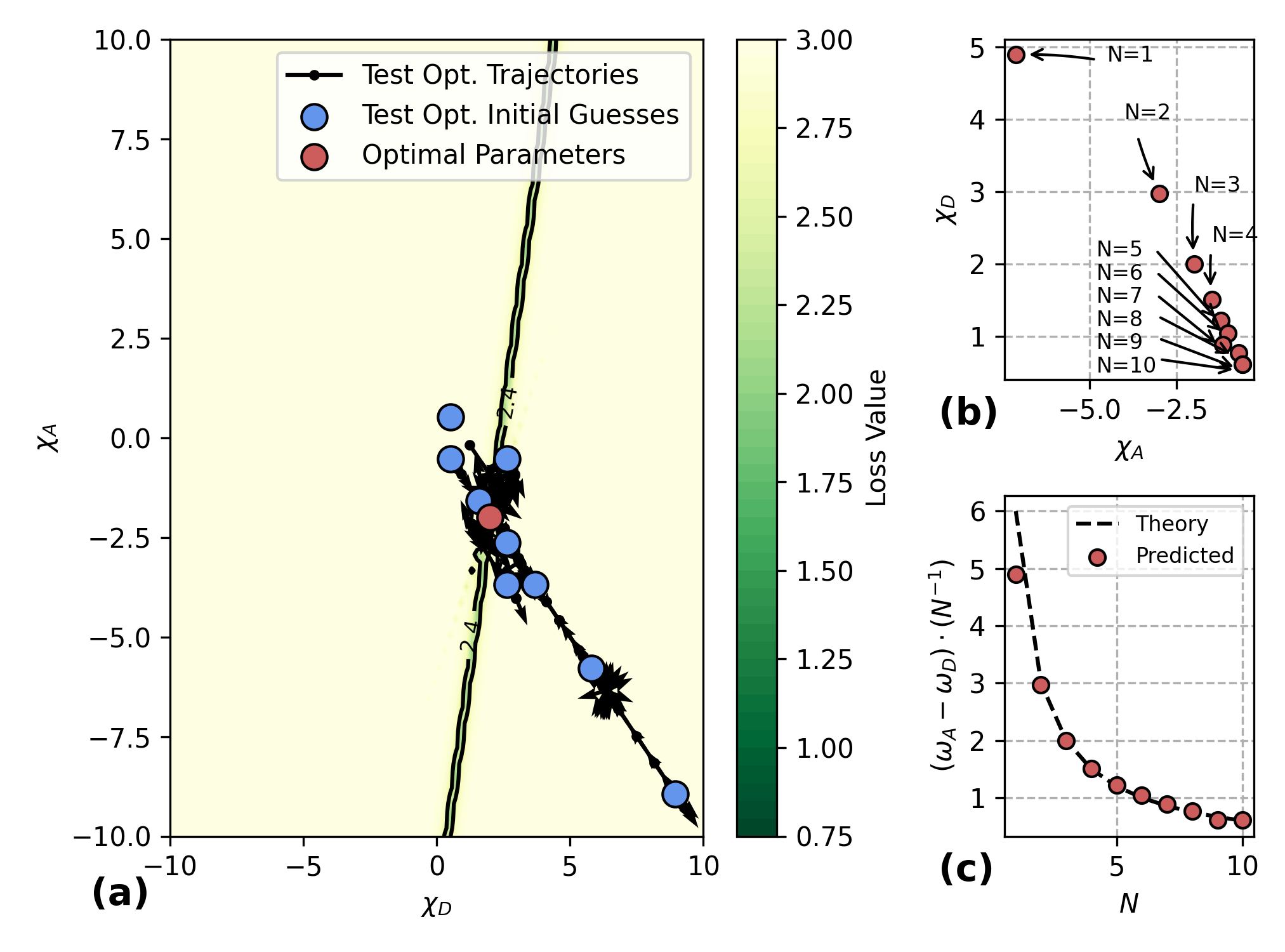}
    \caption{Optimization Results. \textbf{(a)} The trajectories and initial guesses of our algorithm, using the grid method, for $N=3$ bosons.\@ The optimal parameters found -by the main optimizer- in this example are ($\chi_A$, $\chi_D$) = (-1.99, 2).\@ Only the initial guesses of the test optimizers that resulted in a LF smaller than 0.5 are displayed.\@ \textbf{(b)} Parameters for TET from the quantum limit to the semiclassical limit, as deduced from the grid method and the main optimizer, in the dimer system.\@ \textbf{(c)} Comparison of the predicted nonlinearity parameters $\chi_D=-\chi_A$, illustrated in the figure as red dots, with the reference value derived from Eq.~\eqref{ManiadisSolution} represented in the figure as a dashed line.\@ The system's constant parameters are ($\omega_A$, $\omega_D$, $\lambda$, \texttt{maxt}) = (3, -3, 0.1, 25).}
    \label{results}
\end{figure*}
Regardless of whether we use the grid or the splitting into regions method, the trainable parameters are updated in the same way for both the test optimizers and the main optimizer.\@ Given the initial guesses of $\{ \chi_k \}$, $k= 1,2,\dots,f$, the nonlinearity parameters after the m$^{th}$ iteration are updated as following:

\begin{equation*}
    \chi_{j}^{(m+1)} = \chi_{j}^{(m)} - \alpha \vec{\nabla}_j (LF),  
\end{equation*}
where the gradient is computed with Tensorflow's $GradientTape$, as discussed in section \ref{Section:LF}. The learning rate $\alpha$ is a positive real number that defines the rate of change at each iteration while moving towards the minimum of the LF.\@ With the new set of nonlinearity parameters, the optimization procedure moves to the next iteration and it will be interrupted either due to slow convergence or because of reaching maximum iterations. 

The threshold parameter of Fig.~\ref{Training} has a dual role. On one hand, it checks whether the nonlinearity parameters of the current iteration yield a LF close to zero, signifying TET. On the other hand, it determines the slow convergence of the algorithm, and its role is to pause the optimization procedure when there is no significant improvement in minimizing the LF. Thus, its value should be small enough (close to zero) to manifest TET, but still nonzero because otherwise, the stochastic optimization procedure we introduce will lead to an infinite loop. The latter problem is also resolved by terminating the optimization procedure after reaching a predefined maximum number of iterations.

The $Python$ code implementing this procedure is located in our GitHub repository \cite{Repository}.

\section{Results}
\subsection{TET Quantum Dimer}
In the preceding section, we describe  the optimization technique, including the possible alternatives.\@ We may now apply this scheme to the dimer realm and test our method from the semiclassical limit to the fully quantum one, using the grid method described earlier.\@ As we discussed in section \ref{Section:Intro}, the optimal parameters for this case are already known for both the semiclassical \cite{Kopidakis2001} and the quantum regime \cite{Maniadis2004}:

\begin{equation}{\label{ManiadisSolution}}
\chi_D = -\chi_A = \frac{\omega_A-\omega_D}{N}.
\end{equation}
Our method is successful in obtaining the TET configurations.\@ We fix the frequencies of the  donor and acceptor to $\omega_D=-3$ and $\omega_A = 3$  respectively, the coupling parameter to $\lambda=0.1$, while the time evolution of Eq.~\eqref{avg_n} for the acceptor is performed until \verb{maxt{ = 25.\@ The Fig.~\ref{results}a displays the outcome of the optimization procedure for $N=3$ bosons, with the optimal nonlinearity parameters being:
\begin{equation*}
    (\chi_D, \chi_A) = (2,-1.99).
\end{equation*}
\@The illustrated test optimizers produce a LF lower than the threshold mentioned in Fig.~\ref{Training}, which is set arbitrarily (in this case 0.5).\@ The results validate the sensitivity of each optimizer to the initial guesses.

We keep the same values for $\omega_A,\omega_D,\lambda$, and apply the grid method for a variety total of bosons in the dimer system.\@ The outcome of the optimization scheme is shown in Fig.~\ref{results}b, while we compare our results with Eq.~\eqref{ManiadisSolution} in Fig.~\ref{results}c.\@ In every case, the proposed method succeeds in identifying the TET paths.\@ It is worth mentioning that we can deduce the same results with the method of  splitting the parameter space.

Moreover, our analysis proves that the fully quantum case of $N=1$ boson is of special interest since TET occurs for a whole set of nonlinearity parameters $\chi_A,\space \chi_D$ instead of a single, very limited configuration.\@ Specifically, we identify this set as 
\begin{equation}{\label{OneBosonParams}}
    \chi_D = \chi_A + 2(\omega_A-\omega_D).
\end{equation}

This result is in agreement with previous analytical calculations.\@ To be more specific, Maniadis et al.\@ in \cite{Maniadis2004} prove that the condition for having TET is for the detuning function to vanish.\@ The latter is defined as the variation of the energy of the oscillators during a transfer:
\begin{equation}\label{Detuning}
    \epsilon = [ H_D(N)+H_A(0) ]-[ H_D(i) + H_A(N-i)].
\end{equation}
In  Eq.~\eqref{Detuning} $i=0,1,\dots,N$ while $H_D$, $H_A$ are the
donor and acceptor parts of the Hamiltonian in Eq.~\eqref{bosehubbard_hamiltonian}.\@ In this case, any nonzero $\lambda$ can raise the degeneracy of the system and complete TET occurs in the limiting of zero coupling.\@ We can easily prove that the detuning function of Eq.~\eqref{Detuning} for one boson  vanishes under the parameters of Eq.~\eqref{OneBosonParams}.


For all the cases where the detuning function vanishes, the Hamiltonian becomes quadratic of the bosons operators:
\begin{equation}{\label{QuadraticHamiltonian}}
    \hat{H} = \hat{H}_D(N) - \lambda(a^\dagger_D a_A +a^\dagger_A a_D).
\end{equation}

\begin{figure}[t!]
\includegraphics[width=8.5cm,height=14.5cm]{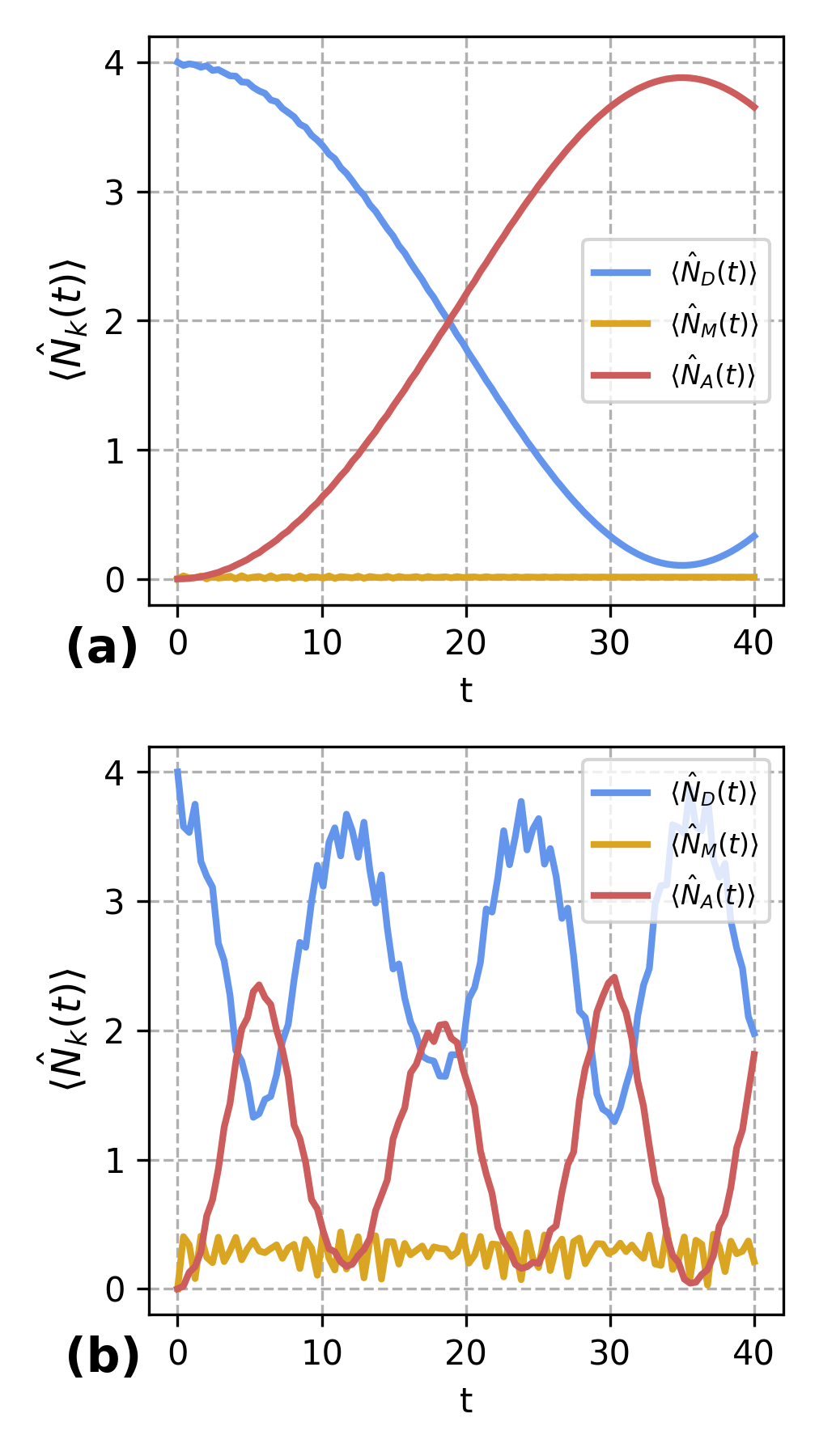}
 \caption{Time evolution of the average number of bosons for the three sites of the trimer system. \textbf{(a)} The time evolution of the expectation values of the boson number operators for nonlinearity parameters that produce near complete TET where ($\chi_A$, $\chi_M$, $\chi_D$, $\omega_D$, $\omega_M$, $\omega_A$, $\lambda$, $N$, \texttt{maxt}) = (-1.5, -38.39, 1.5, -3, -3, 3, 1, 4, 40). \textbf{(b)} The time evolution of the expectation values of the boson number operators for a system with parameters that don't produce complete TET, where ($\chi_A$, $\chi_M$, $\chi_D$, $\omega_D$, $\omega_M$, $\omega_A$, $\lambda$, $N$, \texttt{maxt}) = (-1.5, 1.5, 1.5, -3, -3, 3, 1, 4, 40). For both figures, the blue, yellow, and red lines refer to the expectation value of the boson number operator for the donor, middle site, and acceptor respectively.} 
 \label{TimeEvolutionPlotTrimer}
\end{figure}

\subsection{ TET Quantum Trimer }

The trimer system has been investigated in \cite{Aubry2003,Barmparis2021} in the context of a single electron or boson.\@ We aim to expand this investigation to the arbitrary case of $N$ bosons. While the method is able to optimize the nonlinearity parameters of all three sites, it proved computationally consuming.\@ To circumvent this, we set the acceptor and donor sites to their dimer values, and optimize the parameter of the middle layer (labeled as \enquote{M}). Similarly to the dimer case, we can set an arbitrary threshold for the LF (LF$<$ 0.2) and begin the optimization process.

We observe that for TET to occur in the quantum realm, the middle layer had to exhibit extremely high nonlinearity. One of the trimer systems examined in this section has the following properties:\@ It is strongly coupled, $\lambda=1$, with frequencies $\omega_{D}=3,\;\omega_{M}=-3,\;\omega_{A}=-3$ and a maximum number of bosons $N=4$ initially at the donor site.

\begin{figure*}[t!]
    \includegraphics[width=16cm,height=9cm]{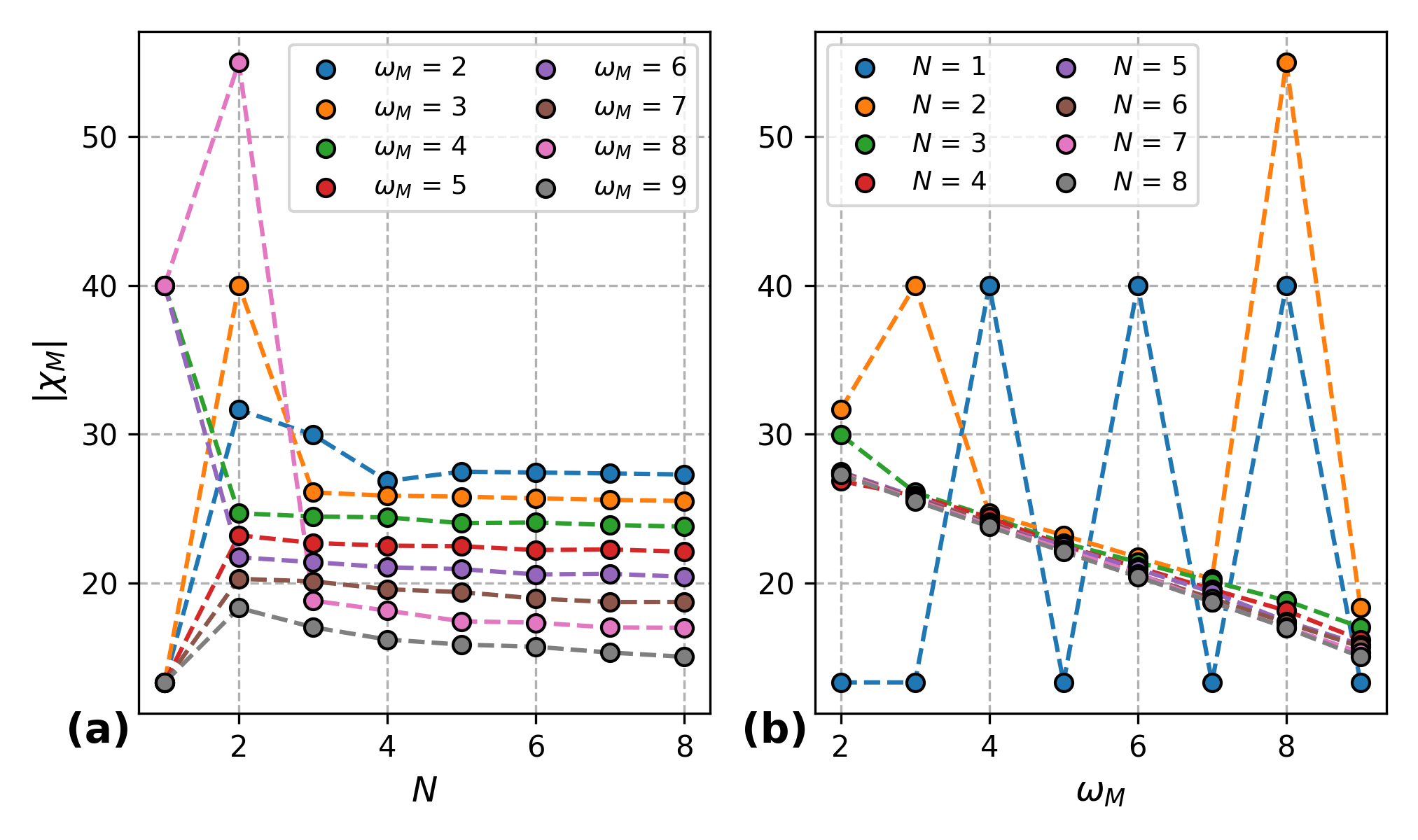}
    \caption{Trimer Results. Absolute values of the nonlinearity parameter of the middle layer with respect to \textbf{(a)} different values of the maximum number of bosons in the system, \textbf{(b)} the frequency of the middle layer $\omega_M$. In both cases we use the grid method while the system's constant parameters for both figures are ($\omega_A$, $\omega_D$, $\lambda$, \texttt{maxt}) = ($\omega_M$-1, -$\omega_M$+1, 1, 40).}\label{results_trimer}. 
\end{figure*}

The nonlinearity parameter required to have TET, in this case,\@ is $\chi_M\approx38.39$,\@ much higher than that of Eq.~\eqref{ManiadisSolution}, compared to the parameters of the donor and acceptor sites $\chi_D=-\chi_A=1.5$.\@ It is important to mention that the opposite value $\chi_M\approx-38.39$ will produce a similar LF value, but still small enough to exceed the threshold and stop the iterative process.\@ We carry out the same procedure for a variety of system parameters, as seen in Fig.~\ref{results_trimer}. 

\begin{figure}[t!]
    \includegraphics[width=7.5cm,height=7.5cm]{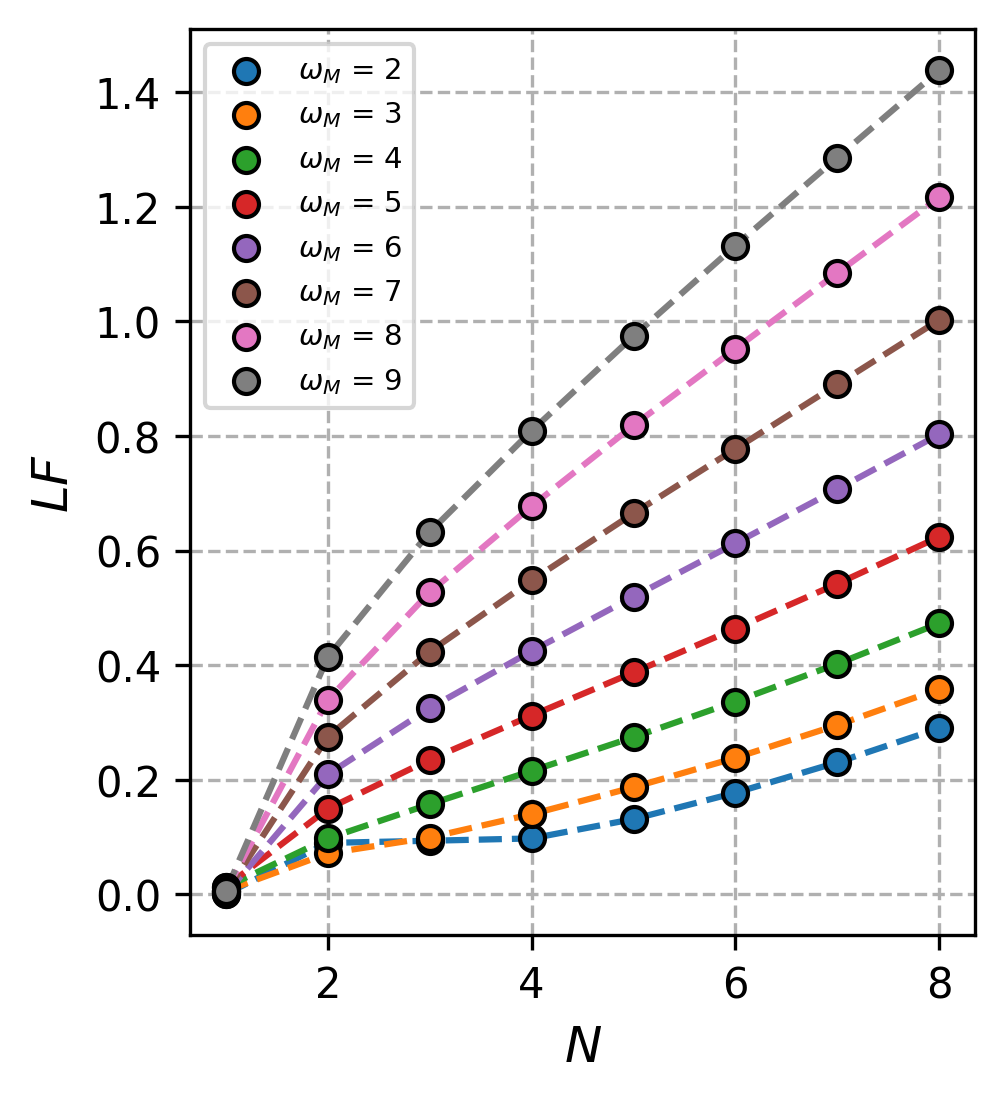}
    \caption{Lowest value of the LF in a trimer system with the parameters displayed on the figure. The other system parameters are ($\omega_A$, $\omega_D$, $\lambda$, \texttt{maxt}) = ($\omega_M$-1, -$\omega_M$+1, 1, 40).}\label{lf_trimer}
\end{figure}

We display the absolute value of $\chi_M$ as both positive and negative values produce the desired result.\@ We observe that the relation between $\chi_M$ and the frequencies or the number of bosons of the system seems to differ from the dimer case, as there appears to be a greater correlation with the former rather than the latter.\@ In the graph some points seem to differ a lot from others, a fact that is attributed to an optimization process that was not able to minimize the loss function enough before being terminated by the rules we defined earlier in section IV. We also have plotted the minimum value of the LF that we observed in Fig.~\ref{lf_trimer}.\@ The time evolution of the boson number operators for this system is shown in Fig.~\ref{TimeEvolutionPlotTrimer}.\@ As we can see in subplot (a) the expectation value of the boson number operator for the middle layer, in the optimal case, appears to be zero for all time steps which indicates that bosons do not stay on this site for any significant amount of time, or at all.\@ However, this is not observed in cases where the nonlinearity parameters are set to nonoptimal values, as seen in subplot (b), where the expectation value is nonzero.\@ It is apparent that the oscillation frequency of the number operator is larger in the non-resonant system and that complete TET cannot be  achieved.

\section{Discussion}

In this work, we introduced a ML method in the context of 
a quantum many-body system and showed that its efficient implementation can produce results that are very hard to obtain with more conventional methods.\@ We focused on the quantized version of the DNLS equation with arbitrary local energies and nonlinearities and addressed the question of optimal transfer in between different sites for the dimer and trimer cases.\@ Since the  fully quantum transfer dimer case is known analytically we compared our method with these results and showed perfect agreement.\@ This successful comparison between analytics and ML methods shows that the latter can be used confidently in more complex cases where results are not known.\@  Subsequently, we applied the method to the trimer case that cannot be 
solved analytically.\@ Our method enabled a detailed search showing the specifics of the resonant transfer,  the different parameter regimes as well as the transfer efficiencies.\@ In terms of physics, we found in the trimer system that in the nonresonant transfer regime from  donor to acceptor sites the intermediate state retains some of the probability.\@ In the resonant case, on the other hand, the intermediate site is essentially not populated.\@  This shows that this site acts as some form of a barrier between the donor and acceptor sites that can be completely bypassed in the fully resonant regime.\@ It is noteworthy to point out that the bosons move in unison over to the acceptor site showing a very interesting collective behavior in the transfer.

The collective boson transfer can be investigated also in more general chains with a larger number of sites.\@ The computational challenge is now larger since the   dimensionality of the system becomes large and the calculation of the Hamiltonian and the evolution of $\langle \hat{N}_A(t) \rangle$ slows down.\@ In this regime, one needs to explore other
loss functions that could improve scalability and/or  implementation of meta-learning methods described in \cite{Li2016}.\@ In this work, we tested also alternative optimizers such as ones with momentum.\@  We found that they were more efficient in finding the resonant transfer parameter regime but were highly dependent on hyper-parameters that needed to be also optimized.\@   Finally,  it
is possible that variants of the gradient descent algorithm might help in reducing the computational cost\cite{Ruder2016}.

The phenomenon of the collective transfer of bosons in the trimer case opens up very interesting  new questions  on the interplay of nonlinearity and disorder in the fully quantum regime for more extended systems.\@ The ML method provided in the present work can be readily generalized to this case and be utilized to investigate this very exciting problem with applications in condensed matter physics as well as quantum optics.
\section*{ACKNOWLEDGMENTS}
We acknowledge the cofinancing of this research by the European Union and Greek national funds through the Operational Program Crete 2020-2024, under the call "Partnerships of Companies with Institutions for Research and Transfer of Knowledge in the Thematic Priorities of RIS3Crete", with project title “Analyzing urban dynamics through monitoring the city magnetic environment” (project KPHP1 - 0029067).
\section*{AUTHOR CONTRIBUTIONS}
I.~Andronis and G.~Arapantonis contributed equally to this work.

\bibliography{bibliography}

\end{document}